\documentclass[aps,floats,prx,showpacs,twocolumn]{revtex4-1}

\usepackage{graphicx,subfigure}
\usepackage{amsfonts,amsmath} \usepackage{bm} \usepackage{dcolumn}
\usepackage{epsfig} \usepackage{latexsym}
\usepackage{graphicx}
\usepackage{multirow}
\usepackage{color}
\usepackage{amsmath}
\usepackage{amssymb}
\def\be{\begin{equation}}
\def\ee{\end{equation}}
\def\bea{\begin{eqnarray}}
\def\eea{\end{eqnarray}}

\def\nn{\nonumber}
\def\half{\frac12}

\begin{document}

\title{Anomalous scalings of the cuprate strange metals from nonlinear\\ electrodynamics}

\author{Sera Cremonini}

\author{Anthony Hoover}

\author{Li Li}

\author{Steven Waskie}

\affiliation{Department of Physics, Lehigh University, Bethlehem, PA, 18018 USA.}

%
%
\begin{abstract}
We examine transport in a holographic model which describes, through a nonlinear gauge field sector, generic nonlinear interactions between the charge carriers.
Scaling exponents are introduced by using geometries which are nonrelativistic and hyperscaling-violating in the infrared.
In the dilute charge limit in which the gauge field sector does not backreact on the geometry,
a particularly simple nonlinear theory reproduces the anomalous temperature dependence of the resistivity and Hall angle of the cuprate strange metals, $R \sim T$ and $\cot\Theta_H \sim T^2$ while also allowing for a linear entropy $S \sim T$, and predicts that the magnetoresistance for small values of the magnetic field $h$ should scale as $\sim h^2 T^{-4}$.
Our study lends evidence to the idea that the strange metal behavior of the cuprates relies crucially on the linear temperature dependence of the entropy.
\end{abstract}

\maketitle

\section{Introduction}
The anomalous metallic state in the high-temperature superconducting cuprates is 
one of the most remarkable puzzles in condensed matter physics~\cite{Keimer:2015}.
The transport properties of the strange metal phase are characterized by universal  
temperature scalings which are robust across widely different systems, and are believed to be controlled 
by an underlying strongly interacting quantum critical sector. 
Magnetotransport measurements have identified  
anomalous power law behavior $\sim T^{n}$ for the in-plane resistivity, the Hall angle and the weak-field magnetoresistance (corresponding to $n=1, 2$ and $-4$, respectively)~\cite{Harris:1995,Hussey:2008,Hartnoll:2015sea}. The universality of these scaling laws, seen in a variety of materials, 
strongly motivates the search for robust physical mechanisms that can reproduce such observations.

The holographic gauge/gravity duality has shown to be a useful theoretical laboratory to probe the physics of strongly interacting quantum critical states, by mapping them onto a dual gravitational problem~\cite{Jan:book,Hartnoll:2016apf}. 
Using the techniques of holography,  in this paper we will study a model that 
describes a sector of probe charge carriers interacting amongst themselves and with a larger set of neutral quantum critical degrees of freedom.
Thus far standard Einstein-Maxwell-dilaton (EMD) theories have been unable to reproduce the 
anomalous scalings of the strange metal phase in the presence of a magnetic field (see e.g. Refs.~\cite{Blake:2014yla,Amoretti:2015gna,Zhou:2015dha,Amoretti:2016cad}). Indeed, there is evidence~\cite{Kim:2010zq,Karch:2014mba,Blauvelt:2017koq} 
that to reliably capture transport in these phases it may be crucial to 
take into account the nontrivial dynamics between the charge degrees of freedom.
This is reasonable given that we are dealing with strongly correlated electron matter, 
and was already suggested in Ref.~\cite{Hartnoll:2009ns}.

With this in mind, we examine a holographic theory which includes a generic nonlinear gauge field sector and compute the associated DC conductivities. 
We consider a particularly simple nonlinear model whose structure is natural from the point of view of the Dirac-Born-Infeld (DBI) action. 
Remarkably, this simple, solvable model 
provides the first holographic realization of 
the temperature scalings of  the entropy $\sim T$, resistivity $\sim T$,  Hall angle $\sim T^2$ and weak-field 
magnetoresistance $\sim T^{-4}$ observed in the cuprates
 -- with a minimal set of assumptions.
The underlying mechanism relies on having a quantum critical IR fixed point 
and on the nonlinear structure of the interactions between the charges.
Our results also suggest that the strange metal behavior 
is intimately tied to the linear temperature dependence of the entropy.

\section{Holographic Setup and Conductivities}
%
We are interested in describing a strongly coupled quantum theory containing a sector of dilute charge carriers that interact amongst themselves as well as with a quantum critical bath. The charge degrees of freedom should be thought of as a probe when compared to the larger set of neutral quantum critical degrees of freedom.
What we have in mind are gravitational theories of the type
\begin{equation}\label{lagrangian}
S = \int d^4x \sqrt{-g} \, \left[\mathcal{L_{\text{Bath}} + L_{\text{U(1)}}} \right] ,
\end{equation}
with a bath sector $\mathcal{L}_{\text{Bath}}$ supported, for example, by a neutral scalar $\phi$ and axionic scalars, 
and a charge sector $\mathcal{L_{\text{U(1)}}}$ describing the dynamics of a $U(1)$ gauge field $A_\mu$. 
In particular, since we are interested in capturing 
generic nonlinear electrodynamics effects, 
the latter will be encoded in the Lagrangian term
$\mathcal{L_{\text{U(1)}}}=\mathcal{L}(s,p,\phi)$, which  
is a generic function of the two combinations
\be
\label{sp}
s = -\frac{1}{4} F_{\mu \nu} F^{\mu \nu} \, , \quad \quad p = -\frac{1}{8} \epsilon_{\mu \nu \rho \sigma} F^{\mu \nu} F^{\rho \sigma} \, ,
\ee
with $F_{\mu\nu}$ the field strength for $A_\mu$ and $\epsilon_{\mu \nu \rho \sigma}$ the covariant Levi-Civita tensor, and allows for  couplings between the gauge field and the neutral scalar $\phi$.
Such theories include as a special case the DBI model studied in Refs.~\cite{Blauvelt:2017koq,Kiritsis:2016cpm} and extend those studied in Ref.~\cite{Guo:2017bru} by adding a scalar sector.

Assuming a background which is homogeneous and isotropic, the quantum critical bath can be described holographically using a black brane geometry of the form
\bea
\label{baremetric}
ds^2 &=& -U(r) \, dt^2 + \frac{dr^2}{U(r)} + C(r) (d x^2+d y^2) \,,
\eea
with a nontrivial scalar $\phi = \phi(r)$ depending on the holographic radial coordinate $r$. 
The holographic DC conductivities associated with the conserved current $J^\mu$ dual to $A_\mu$
can be obtained following the prescription developed by Ref.~\cite{Iqbal:2008by} (see also Ref.~\cite{Guo:2017bru}).
In the probe limit the DC conductivity matrix $\sigma_{ij}$ for the broad class of theories (\ref{lagrangian})
is only sensitive to the structure of $ \mathcal{L}(s,p,\phi) $ and in particular is given by
\bea
\label{sigmasprobe}
\sigma_{xx} &=& \mathcal{L}^{(1,0,0)} \, , \quad \sigma_{xy} = -\mathcal{L}^{(0,1,0)}  \, , 
\eea 
where we have defined for convenience
\be
\mathcal{L}^{(1,0,0)}  \equiv \frac{\partial \mathcal{L}(s,p,\phi)}{ \partial s} \, , \qquad  \mathcal{L}^{(0,1,0)} \equiv \frac{\partial \mathcal{L}(s,p,\phi)}{ \partial p} \, .
\ee
The corresponding resistivity and Hall angle are then
\bea
\label{RThetaprobe}
R_{xx} = \frac{\sigma_{xx}}{\sigma_{xx}^2+\sigma_{xy}^2} &=& \frac{\mathcal{L}^{(1,0,0)} }{\left(\mathcal{L}^{(1,0,0)}\right)^2 + \left(\mathcal{L}^{(0,1,0)}\right)^2 } \, ,   \nn \\
\cot \Theta_H = \frac{\sigma_{xx}}{\sigma_{xy}} &=& -\frac{\mathcal{L}^{(1,0,0)}}{\mathcal{L}^{(0,1,0)}} \,, 
\eea 
where it should be understood that all functions are evaluated at the horizon $r=r_h$ of the black brane (\ref{baremetric}) 
whose temperature is $T =U^\prime(r_h)/4\pi$. 

Note that when $\mathcal{L}^{(0,1,0)}=0$, $\sigma_{xy}=0$  and hence $\tan \Theta_H=0$.
Thus, the presence of $p\sim F \wedge F$ in the theory leads to a distinctively different behavior for the conductivities.
As an example,  in the standard  linear EMD theory 
$\mathcal{L}(s,p,\phi)=Z(\phi) s$ thus far it has been difficult  
to realize the scaling behavior of the cuprates.
In the probe limit this situation is not ameliorated, because although 
the associated resistivity $ R_{xx} = 1/Z$ 
can in principle be engineered to be linear, the Hall conductivity is trivial.
This compels us to study nonlinear electrodynamics effects.

Finally, we stress that the result (\ref{RThetaprobe}) is quite general, as it relies only on a minimal set of assumptions -- a homogeneous and isotropic metric
(\ref{baremetric}) modeling the quantum critical bath, and the presence of a dilute set of charge carriers.
The analysis of the DC conductivities away from the probe limit is included in the Supplemental Material~\cite{newsm}, 
where it can be seen that $\sigma_{ij}$ in the backreacted case is sensitive not only to the gauge field sector, but also to the geometry and the structure of the model supporting the quantum critical bath. 
Interestingly, we find that the probe limit results can be obtained from the fully backreacted case when the scale of momentum dissipation dominates over the other physical scales in the system.

\section{Quantum Critical Bath Geometry}
%
Motivated by condensed matter studies of quantum criticality in strange metals~\cite{Fitzpatrick:2013mja,Fitzpatrick:2013rfa,Hartnoll:2014gba}, 
we will be specifically interested in solutions that are nonrelativistic and violate hyperscaling in the IR of the geometry -- thus, the dual system will be quantum critical in a generalized sense.
To work with the standard holographic dictionary we consider geometries which asymptote to anti-de Sitter (AdS) at the boundary.
The IR scaling behavior of such geometries will lead naturally to clean scaling regimes in the holographic transport coefficients and in 
particular in the DC conductivities, which are determined by horizon data.

\begin{figure}
\begin{center}
\includegraphics[width=3.1in]{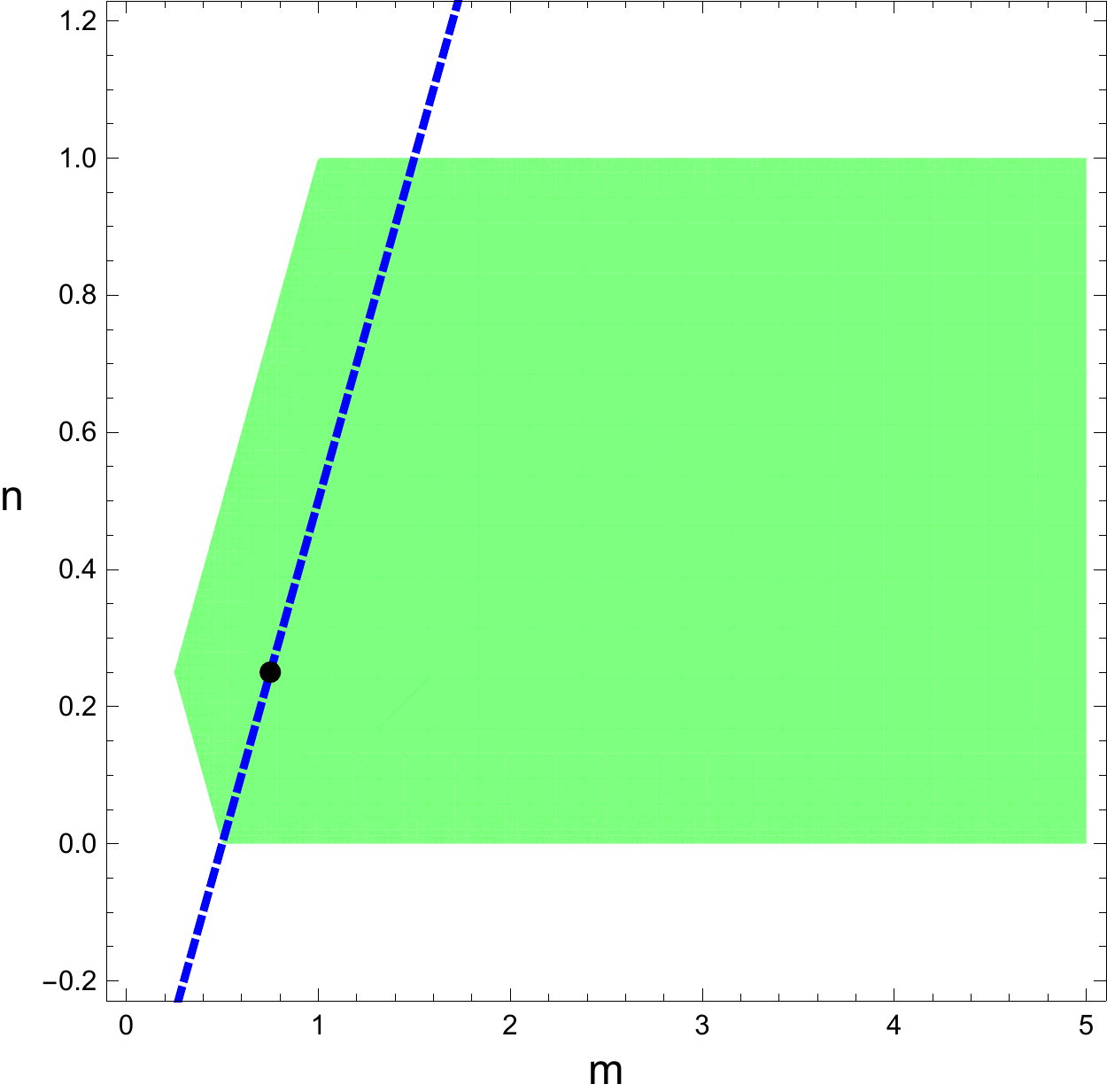}
\caption{The green region denotes the values of $(m,n)$ that satisfy the physical constraints which ensure a well-defined ground state geometry. 
The dashed blue line shows the case which gives a linear entropy $S\sim T$. The black dot corresponds to the $\eta$-geometry with $m=3/4, n=1/4$ (or $\eta=1$).}
\label{figmn}
\end{center}
\end{figure}

One advantage of working in the probe limit is that 
we have a clean separation between the background geometry and the gauge field sector.
In particular, a simple holographic model which supports analytical scaling geometries 
contains a dilatonic scalar $\phi$ and two axionic scalars $\psi^I$,
\begin{equation}
\mathcal{L}_{\text{Bath}} =\mathcal{R} - \dfrac{ (\partial \phi)^2}{2} - V(\phi) 
-  \dfrac{Y(\phi)}{2} \sum_{I=1}^{2} (\partial \psi^I)^2\,.
\end{equation}
When the scalar couplings are well approximated by single exponentials in the IR of the geometry,
\begin{equation}
Y= e^{\alpha \phi} \, , \qquad V=-V_0 e^{-\beta\phi}  \, ,
\end{equation}
with $\alpha, \beta$ and $V_0$ constants, the theory supports 
the following hyperscaling violating, Lifshitz-like black-branes,
\bea
\label{mnsols}
&& ds^2 = -\left(\frac{r}{\ell}\right)^{2m} f(r) dt^2 + \left(\frac{r}{\ell}\right)^{-2m} f(r)^{-1} dr^2 + \left(\frac{r}{\ell}\right)^{2n} d\vec{x}^2, \nn \\
&& f(r)= 1- \left( \frac{r_h}{r} \right)^{2m+2n-1}\,, \quad \phi(r) =\kappa \ln \left(\frac{r}{\ell}\right) \, ,    \nn \\
&& \kappa^2 = 4n(1-n), \;\; \alpha \kappa = 2(m+n-1), \;\; \beta \kappa = 2(1-m),\nn \\
&& \ell^{2} V_0 = 2m(2m+2n-1)  \, , \quad  k^2 = \frac{(m-n) V_0}{m} \, , 
\eea
where we have chosen the axion configuration 
\begin{equation}
\psi^1=k\, x,\quad \psi^2=k\,y\,,
\end{equation}
with the constant $k$ denoting the strength of momentum dissipation. 
The temperature of these solutions scales with the horizon radius as $T\sim r_h^{2m-1}$ and the entropy as 
$S \sim T^{\frac{2n}{2m-1}}$. 
The scaling parameters $n,m$ can be related to the standard dynamical critical exponent $z$ and hyperscaling violating exponent $\theta$ by using
\be
m=\half \frac{\theta-2z}{\theta-z} \, , \qquad  n =\half \frac{\theta-2}{\theta-z} \, .
\ee
In terms of $n, m$ the $\eta$-geometries discussed in Ref.~\cite{Hartnoll:2012wm}, which arise from taking the 
limit $z \rightarrow \infty, \theta \rightarrow \infty$ 
with $\eta \equiv - \theta/z$ held fixed, correspond to taking $n+m=1$ (with $\eta = \frac{2n}{1-2n} $). 
Finally, since these scaling geometries are generically singular, one needs to impose appropriate constraints on the parameter space in order to ensure a well-defined ground state solution, including Gubser's criterion~\cite{Gubser:2000nd} and the null energy condition (see e.g. the discussion in Refs.~\cite{Charmousis:2010zz,Cremonini:2016bqw}). 
The parameter space corresponding to physically acceptable ranges for the exponents $(m,n)$ is shown in Fig.~\ref{figmn}, with the dashed blue line denoting values for which the entropy is linear with temperature.
%

\section{A Specific Nonlinear Model}
Motivated in part by the DBI analysis~\cite{Hartnoll:2009ns,Blauvelt:2017koq}, we choose to focus on a particularly simple nonlinear model,
\be
\label{ourtheory}
\mathcal{L}(s,p,\phi)  = Z(\phi) s + \half Z^2(\phi) p^2 \, ,
\ee
characterized by a single scalar coupling $Z(\phi)$. 
The combination (\ref{ourtheory}) is natural from the viewpoint of the DBI action
\begin{equation}
\begin{split}
S_{DBI}&=\sqrt{-g}-\sqrt{-\det(g+Z^{1/2}\, F)}\\
            &=\sqrt{-g}\left[1-\sqrt{1-2(Z s + Z^2 p^2/2})\right]\,,
\end{split}
\end{equation}
and, despite its simplicity, will turn out to be sufficient to describe four of the scalings observed in the strange metal phase of the cuprates.
Evaluating (\ref{sigmasprobe}) for our nonlinear model (\ref{ourtheory}), 
we find that the resistivity and Hall angle are 
\begin{eqnarray}
\label{Rcot1}
&& R_{xx} \sim \frac{Z}{Z^2 +  Z^4   \, p^2 } = \frac{(C^2 + h^2 Z)^2 }{Z\left[(C^2  + h^2 Z)^2 +  h^2 \rho^2 \right] }   \, , \nn \\ 
&& \cot \Theta_H \sim   -\frac{1}{Z \, p } = \frac{ C^2  + h^2 Z }{  h \rho} \, ,
\end{eqnarray}
with the charge density
\begin{eqnarray}
\rho = \dfrac{A_t' \, Z}{C} \bigg(C^2  + h^2 Z \bigg)\,,
\end{eqnarray}
and all functions evaluated at the horizon.
The expressions (\ref{Rcot1}) can be simplified further by taking a small $h$ limit, 
which is consistent with our assumption of a dilute charge sector. 
This yields the simple expressions
\begin{eqnarray}
\label{Rcot}
R_{xx} \sim \frac{1}{Z}  \, , \qquad \cot \Theta_H \sim  \dfrac{C^2}{ h  \rho} \, ,
\end{eqnarray}
each controlled by a different scale, the first by the scalar coupling $Z$ and the second by the geometry through $C$.
Finally, working with the small $h$ expansion of the resistivity (\ref{Rcot1}) 
we compute the magnetoresistance, 
\be
\label{MRexp}
MR = \frac{R_{xx}(h) - R_{xx}(h=0)}{R_{xx}(h=0)} \sim -  \frac{h^2  \rho^2}{ C^4} \, .
\ee
The results (\ref{Rcot}) agree with those obtained in the DBI construction of Ref.~\cite{Blauvelt:2017koq}, for small values of the charge density and magnetic field. 
While this agreement is expected, since (\ref{ourtheory}) is part of the low-energy expansion of the DBI model, it also suggests that the much simpler nonlinear model (\ref{Rcot}) may suffice to capture the key physics of more complex DBI-like theories. 
In addition, the particular DBI model of Ref.~\cite{Blauvelt:2017koq} predicts a weak-field magnetoresistance that goes as $h^2/T^3$, instead of the $h^2/T^4$ behavior of the cuprates, and therefore it may be more appropriate 
to describe other strange metal phases.

\section{Physical Implication}
We are now ready to comment on the implications of our results, and ask in particular whether the transport quantities we computed can describe the scaling behaviors observed in the cuprates.
First of all, note that while $R_{xx}$ depends on the scalar coupling $Z$, the Hall angle and magnetoresistance (at small $h$) 
are both controlled by the metric component $C$, which also determines the thermodynamic entropy $S$ of the dual field theory,
$S \sim C$.
While the coupling $Z$ can be chosen freely without affecting the geometry,
the function $C$ is fixed for a given background.
Once the geometry is specified -- and therefore the temperature dependence of the entropy -- there is very little freedom in the system.

In our model a linear resistivity $R_{xx}~\sim T $ can be realized by identifying a clean temperature scaling regime for the scalar coupling, of the form
\be
\label{ZCondition}
Z \sim \frac{1}{T}  \, .
\ee
Moreover, experimental data on the cuprates~\cite{Loram:1993,Loram:2001} indicates that the entropy is linear in temperature,
\be
\label{linearS}
S \sim T \, ,
\ee
which requires the spatial metric component to scale as
\be
\label{Ccond}
C \sim T \, ,
\ee 
and unambiguously fixes the temperature dependence of the Hall angle and magnetoresistance in our model to be
\be
\label{RThetaMR}
\cot \Theta_H \sim T^2 \, , \;\; MR \sim - \frac{h^2}{T^4} \, .
\ee 

In our setup (\ref{ZCondition}) and (\ref{Ccond}) can indeed be realized quite naturally, making use of  the quantum critical geometry.
In particular, (\ref{ZCondition}) can be obtained by making
the simple single exponential choice  $Z(\phi)~\sim  e^{\gamma\phi}$ with $\gamma = \frac{1-2m}{2-2m} \beta$.
Moreover, the entropy associated with (\ref{mnsols}) is given by
\begin{equation}
S \sim C(r_h) \sim T^{\frac{2n}{2m-1}} \, .
\end{equation}
Thus, we have a linear entropy $S\sim T$ when $2n = 2m-1$. For $z$ and $\theta$ finite this translates to the condition 
$z = 2-\theta $ corresponding to a one-parameter family of black brane solutions, while for the case of $\eta$-geometries for which both exponents are infinite  we have $m+n=1$ and thus $n=1/4$ and $m=3/4$.
The parameter choices that correspond to a linear entropy are represented by the dashed line in Fig.~\ref{figmn}, with the dot denoting the special case corresponding to
the $\eta=1$ geometry.

It is intriguing and unexpected that the choice (\ref{ZCondition}) and the experimental observation (\ref{linearS})  are sufficient to reproduce the observed
scaling properties of the cuprates. In particular, what we have seen is that the simple nonlinear model (\ref{ourtheory}) 
supports the following behaviors,
\begin{equation}
S \sim T,  \;\;  R_{xx} \sim T\, , \;
\cot \Theta_H \sim T^2 \, , \; MR \sim - \frac{h^2}{T^4} \, . 
\end{equation}
It is convenient to rescale the temperature and magnetic field and work with 
dimensionless quantities. In particular, if we introduce two positive constants $z_0$ and $c_0$ through $Z=z_0/T$ and $C=c_0 T$ (the values of the constants depend on the specific theory one examines), we can construct the dimensionless expressions
\begin{equation}\label{rescale}
\textbf{T}=\frac{c_0^2 z_0}{\rho^2}T,\quad \textbf{h}=\frac{c_0^2 z_0^2}{\rho^3}h\,.
\end{equation}
We then have 
\begin{equation} 
\begin{split}
\label{Dimensionless}
R_{xx}&=\zeta_0\, \textbf{T}    \left[  1+ \frac{\textbf{T}^2\textbf{h}^2}{ \left(\textbf{T}^3+\textbf{h}^2\right)^{2}}  \right]^{-1}, 
 \;\; \zeta_0 \equiv \frac{\rho^2}{c_0^2 z_0^2}\,,\\
\cot \Theta_H&=\frac{\textbf{T}^2}{\textbf{h}}\left(1+\frac{\textbf{h}^2}{\textbf{T}^3}\right)\,,\\
MR&=-\frac{\textbf{T}^2\textbf{h}^2}{(\textbf{T}^3+\textbf{h}^2)^2+\textbf{T}^2\textbf{h}^2}\,.
\end{split}
\end{equation}
We immediately observe that $R_{xx}\sim \textbf{T}$ approaches zero as $\textbf{T}\rightarrow 0$, and in particular $R_{xx}=\zeta_0 \textbf{T}$ in the absence of a magnetic field. Thus, this system indeed describes a metal phase. 

\begin{figure}
\begin{center}
\includegraphics[width=3.0in]{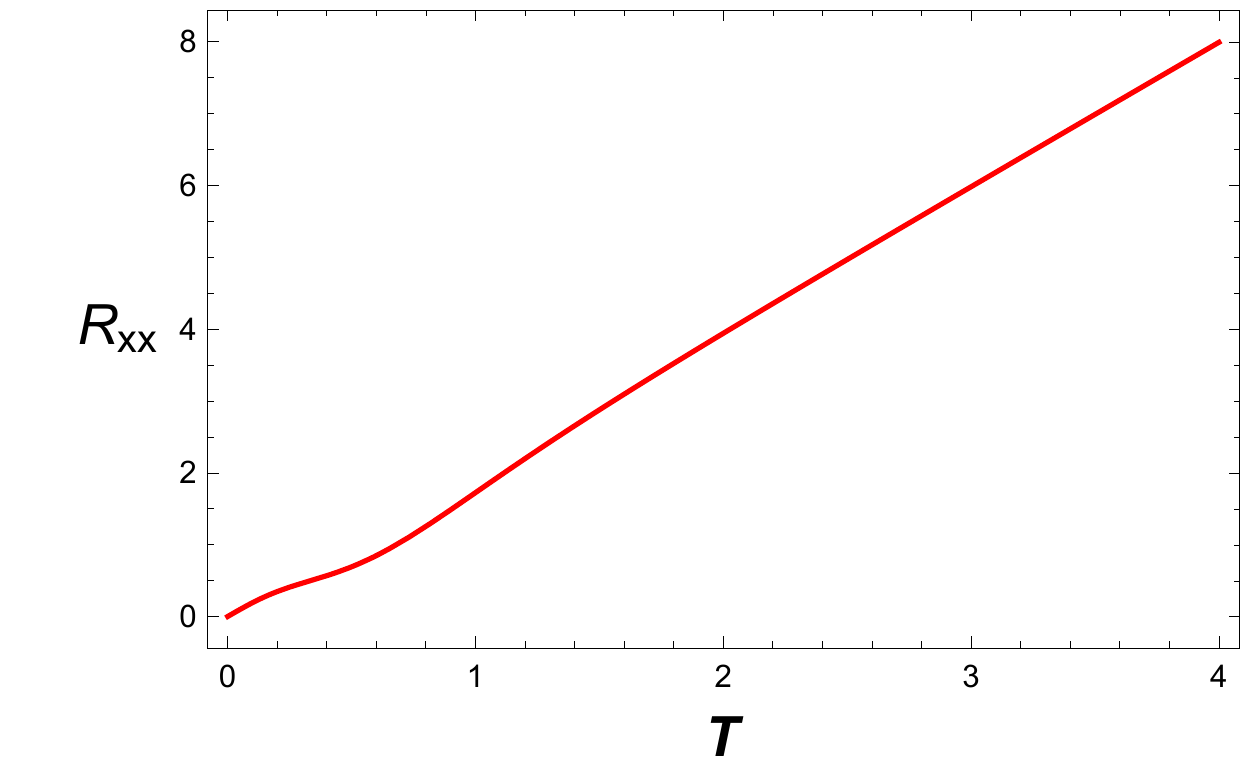}
\includegraphics[width=3.05in]{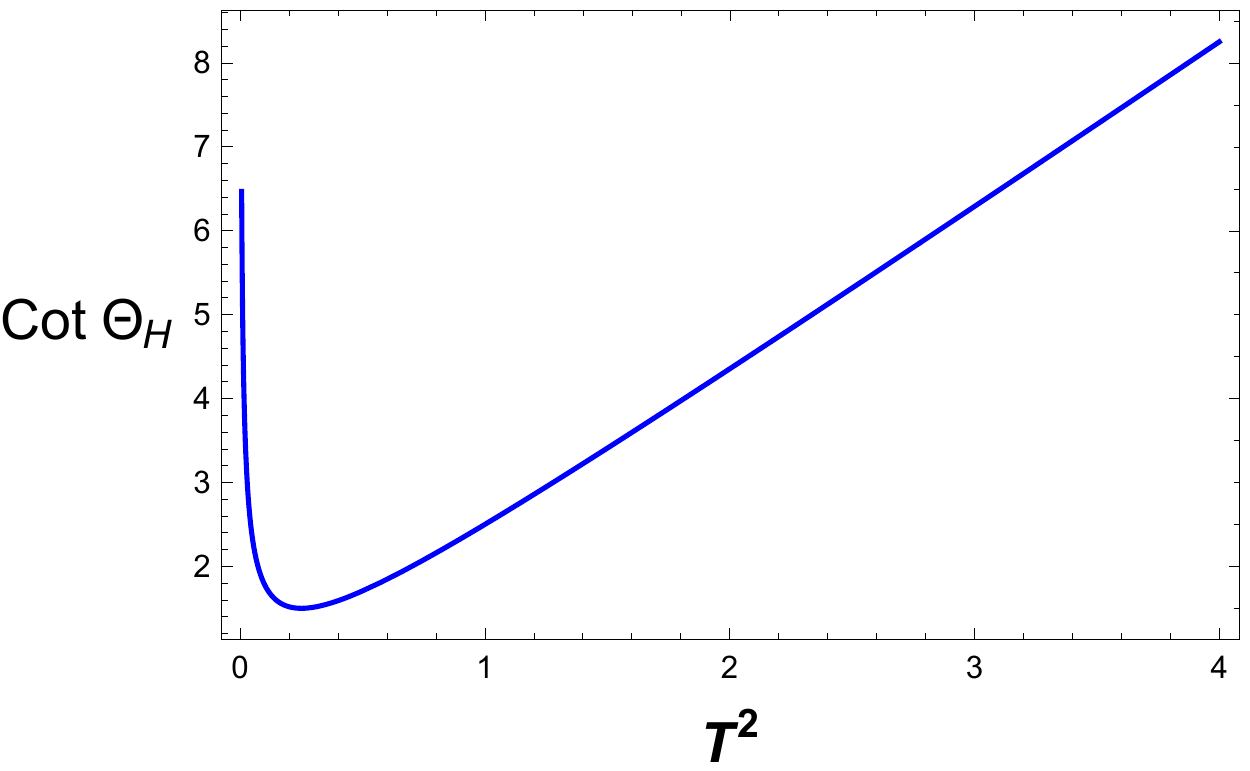}
\includegraphics[width=2.98in]{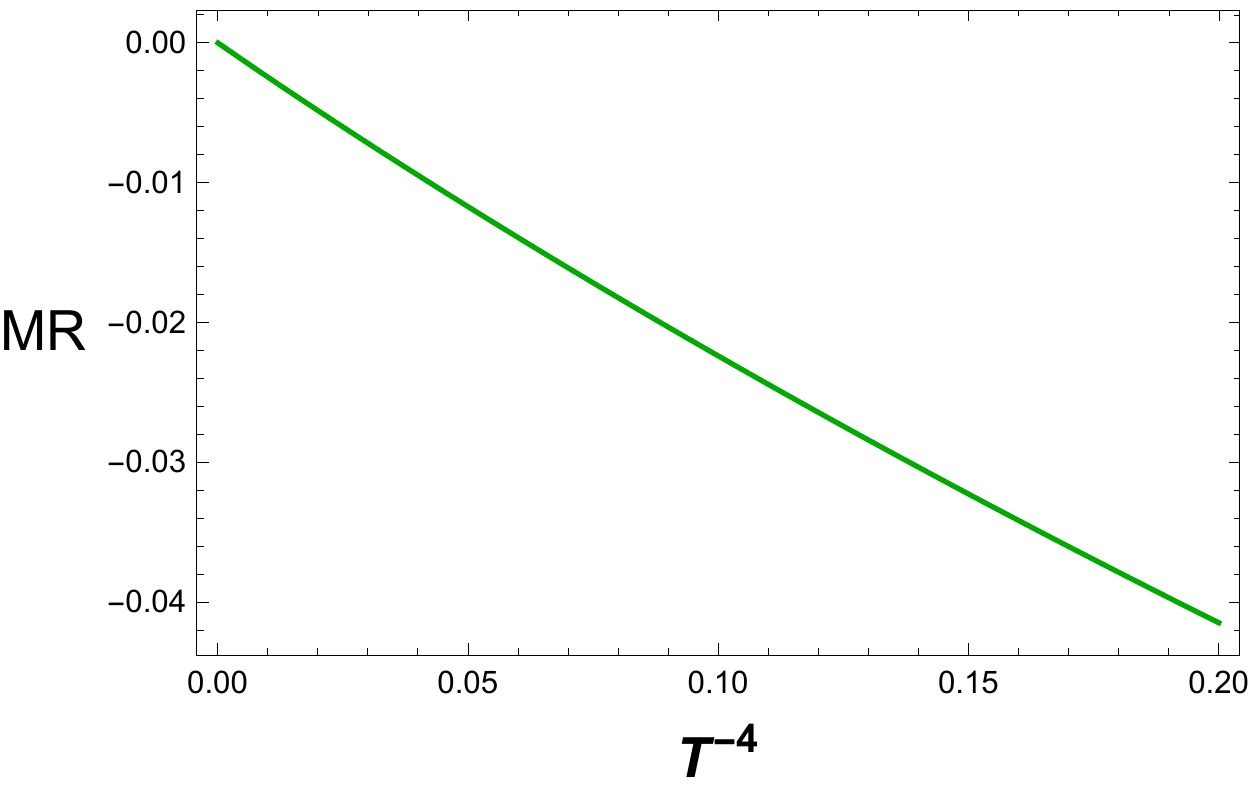}
\caption{Dependence of the resistivity, Hall angle and magnetoresistance in the expressions (\ref{Dimensionless}) on the 
dimensionless temperature $\textbf{T}$. We fix $\textbf{h}=0.5$ and find good scaling behaviors when $\textbf{T}\gtrsim 3 \textbf{h}=1.5$. We have chosen $\zeta_0=2$. }
\label{figmetal}
\end{center}
\end{figure}

The temperature dependence of the quantities (\ref{Dimensionless}) is plotted in 
Fig.~\ref{figmetal}, from which it is clear that when the value of $\textbf{T}$ is sufficiently bigger than $\textbf{h}$
(more precisely, when $\textbf{T}^3 >>\textbf{h}^2$), one realizes the strange metal scalings
\begin{equation}
\label{DimlessScalings}
S\sim \textbf{T}, \; R_{xx}\sim\zeta_0 \textbf{T}, \; \cot \Theta_H\sim\frac{\textbf{T}^2}{\textbf{h}}, \; MR\sim -\frac{\textbf{h}^2}{\textbf{T}^4}.
\end{equation}
Our discussion is based on the rescaled temperature $\textbf{T}$ which is defined with respect to the scale $\rho^2/(c_0^2 z_0^2)$, as seen from~\eqref{rescale}. 
Therefore, note that the regime we are considering is not necessarily a high$-T$ limit.
It could indeed describe low temperature physics provided that this scale is sufficiently higher than the temperature the experiment is probing.

In closing, we would like to mention a few additional features that are visible in our analysis.
First, note that in the absence of a magnetic field the linear temperature dependence of the resistivity is exact.
Moreover, as long as we are away from the transition regime in which $\textbf{T}$ and $\textbf{h}$ are comparable, the 
in-plane resistivity (\ref{Dimensionless}) is not very sensitive to the magnetic field.
As a result, in the temperature range in which one can realize (\ref{DimlessScalings}), the magnetoresistance is small (and negative), 
suggesting that the effect of a magnetic field does not alter the underlying normal state. This is consistent with the observation in the experiment of Ref.~\cite{Daou:2008}.
Finally, we note that the increase in the Hall angle at very small temperatures (visible in Fig.~\ref{figmetal}) is 
similar to the behavior observed in Refs.~\cite{Rosenbaum:1998,Konstantinovic:2000}.

\section{Final Remarks}
Our analysis has provided the first holographic realization of the temperature scalings  
(\ref{DimlessScalings}) through a simple and rather minimal nonlinear model.
In particular, our results relied crucially on the presence of nonlinear interactions among the charge carriers,
which appear to be a necessary ingredient for a bulk description of strongly correlated electron matter.
Working in the dilute charge limit allowed us to construct 
a neutral quantum critical bath, key to identifying clean scaling regimes in the transport coefficients.

In the model (\ref{ourtheory}) the  scaling of the Hall angle and of the magnetoresistance, $T^2$ and  $T^{-4}$ respectively, were 
entirely fixed by the observation that the entropy in the cuprates should be linear in temperature~\cite{Loram:1993,Loram:2001}, $S \sim T$.
Note that the linear resistivity $R_{xx} \sim T$ 
would then  follow immediately by requiring $R_{xx}$ to scale like the entropy, making the choice (\ref{ZCondition}) natural.
Although~\cite{Davison:2013txa,Zaanen:2018edk} argues that the resistivity should indeed be proportional to the entropy, 
their arguments are not applicable in our model -- the inclusion of axions leads to a temperature dependent 
shear viscosity to entropy ratio~\cite{Hartnoll:2016tri}. 
Nonetheless, our study lends evidence to the idea that the cuprates' strange metal behavior depends crucially on the linear  entropy as well as on the existence of a strongly coupled quantum critical IR sector. 

Note that the magnetoresistance associated with  (\ref{ourtheory}) is negative.
Possible mechanisms for a negative magnetoresistance have been proposed for example in Refs.~\cite{Fournier:2000,Sekitani:2003,Hayes:2018nr}.
We should stress, however, that both the sign and scaling properties
of quantities such as the magnetoresistance depend
entirely on the specific model one works with 
(as an example we refer the reader to the DBI analysis in the Supplemental Material~\cite{newsm}). 
Indeed, the minimal model (\ref{ourtheory})  is only one among a large class of nonlinear  
theories one could construct, which generically lead to a highly nontrivial transport structure, as shown in~\eqref{RThetaprobe} (and in (S26) away from the probe limit in the Supplemental Material~\cite{newsm}).
Understanding this rich structure in more detail is 
especially important given the additional scaling regimes that have been recently observed in different high-T$_c$ superconductors, including 
the cuprates \cite{Gallo:2017} and also iron pnictides~\cite{Hayes:2016}. 

With this in mind, 
we hope that our results can provide guidance towards the construction of more realistic theories and help build intuition for the mechanisms 
underlying the unconventional behavior of the cuprates. Since the bulk nonlinearities are related to OPE coefficients of the dual CFT, it would be interesting if the bulk analysis could be used to constrain the structure of the dual field theory and provide physical intuition.
The question remains to what extent such nonlinear holographic models can capture the universality of the strange metal phase. It would be desirable to have a more basic understanding of the role of nonlinear effects on transport. For example, it would be valuable to identify physical regimes in which additional nonlinear terms (coming e.g. from further expanding the DBI action) are subdominant and can be neglected. That would serve to better motivate the minimal model we have studied in this paper.

\onecolumngrid
\vspace{0.5in}
\qquad\qquad\qquad\qquad\qquad\qquad\qquad\qquad\qquad{\textbf{ACKNOWLEDGEMENT}}
\vspace{0.2in}

We would like to thank Sean Hartnoll, Chris Pope, Mike Blake and Richard Davison for valuable conversations.
The work of S.C., A.H. and S.W. is supported in part by the National Science Foundation Grant PHY-1620169.

\twocolumngrid

\pagebreak
\onecolumngrid

\begin{center}
\textbf{\large Supplemental Material}
\end{center}
\setcounter{equation}{0}
\setcounter{figure}{0}
\setcounter{table}{0}
\setcounter{page}{1}
\makeatletter
\renewcommand{\theequation}{S\arabic{equation}}
\renewcommand{\thefigure}{S\arabic{figure}}
\renewcommand{\bibnumfmt}[1]{[S#1]}
\renewcommand{\citenumfont}[1]{S#1}

\qquad\qquad\qquad\qquad\qquad\quad{\textbf{Holographic Conductivity with Generic Gauge Sector}}
\vspace{0.1in}

The holographic theory we examine describes gravity coupled to a neutral scalar field $\phi$, two axions $\psi^I$$(I=1,2)$ and a $U(1)$ vector field $A_\mu$,
\be
\label{action}
S = \int d^4x \sqrt{-g} \bigg(\mathcal{R}  - \dfrac{1}{2} (\partial \phi)^2 - V(\phi) - \dfrac{Y(\phi)}{2} \sum_{I=1}^{2} (\partial \psi^I)^2 - \mathcal{L}(s,p,\phi) \bigg),
\ee
where the term $\mathcal{L}(s,p,\phi)$ describes the gauge field dynamics and is a generic function of the two combinations
\be
s = -\frac{1}{4} F_{\mu \nu} F^{\mu \nu} \, , \quad \quad p = -\frac{1}{8} \epsilon_{\mu \nu \rho \sigma} F^{\mu \nu} F^{\rho \sigma} \, ,
\ee
with $F_{\mu\nu}=\partial_\mu A_\nu-\partial_\nu A_\mu$  and $\epsilon_{\mu \nu \rho \sigma}$ the covariant Levi-Civita tensor. 
A natural constraint is the requirement that in the weak flux limit $F\rightarrow 0$ one should recover the standard gauge field kinetic term, $\mathcal{L}(s,p,\phi)\sim s$. 
The model allows for couplings between the gauge field combinations $(s,p)$ and the scalar $\phi$ and encodes generic nonlinear electrodynamics effects, such as DBI-like interactions arising in string theory. 
In particular, it describes the DBI-based model studied in~\cite{SCremonini:2017qwq} and generalizes the analysis of~\cite{SWang:2018hwg} by introducing scalar couplings.

The equations of motion associated with (\ref{action}) are
\begin{eqnarray}
&& \mathcal{R_{\mu \nu}} - \frac{1}{2} \mathcal{R}g_{\mu \nu} = \frac{Y(\phi)}{2} \sum_{I=1}^{2} \partial_{\mu} \psi^I \, \partial_{\nu} \psi^I + \frac{1}{2} \partial_{\mu} \phi \, \partial_{\nu} \phi - \frac{1}{2} \frac{\partial \mathcal{L}(s,p,\phi)}{\partial s} F_{\mu}^{\phantom{\mu} \sigma} F_{\nu \sigma}\label{einstein}\nn  \\
&&\qquad\qquad\qquad\quad - \half g_{\mu \nu} \bigg( \frac{1}{2} (\partial \phi)^2 +  \frac{Y(\phi)}{2} \sum_{I=1}^{2} (\partial \psi^I)^2 +  V(\phi) + \mathcal{L}(s,p,\phi) - p \frac{\partial \mathcal{L}(s,p,\phi)}{\partial p} \bigg)\, , \\
&& \nabla_{\mu} \nabla^{\mu} \phi - \partial_{\phi} V(\phi) - \half \partial_{\phi} Y(\phi) \sum_{I=1}^{2} (\partial \psi^I)^2 - \partial_{\phi} \mathcal{L}(s,p,\phi) = 0\,,   \\
&& \nabla_{\mu} (Y(\phi) \nabla^{\mu} \psi^I) = 0\,,  \\
&& \nabla_{\mu} G^{\mu \nu} = 0\label{maxwell}\,,
\end{eqnarray}
where the tensor $G_{\mu \nu} $ is given by
\begin{eqnarray}
G^{\mu \nu} = \frac{\partial \mathcal{L}(s,p,\phi)}{\partial s} F^{\mu \nu} + \frac{1}{2} \frac{\partial \mathcal{L}(s,p,\phi)}{\partial p} \epsilon^{\mu \nu \rho \sigma} F_{\rho \sigma}\,.
\end{eqnarray}
We are interested in finite temperature solutions to this theory which approach $AdS$ at the boundary.
Assuming homogeneity and isotropy, we consider the bulk metric and the matter fields of the background geometry to be of the following generic form,
\begin{equation}
\begin{split}
\label{bk}
ds^2 =& -U(r) \, dt^2 + \frac{dr^2}{U(r)} + C(r) (d x^2+d y^2)\,, \\
\phi=& \phi(r), \hspace{.5cm} \psi^1 = k\,x, \hspace{.5cm} \psi^2 = k\,y, \hspace{.5cm} A = A_t(r) dt + \frac{h}{2}(x\, dy - y\, dx)\,,
\end{split}
\end{equation}
with $h$ denoting the magnitude of the magnetic field. The temperature associated with these black branes is given by  
\begin{eqnarray}\label{temp}
T = \frac{U'(r_h)}{4 \pi}\,,
\end{eqnarray}
with $r_h$ denoting the horizon radius. 
The linear dependence of the axionic scalars $\psi^I$ on the spatial coordinates of the boundary field theory breaks translational invariance giving rise to momentum dissipation, whose strength is parametrized by $k$.
Finally, from Maxwell's equation~\eqref{maxwell} we obtain a radially independent quantity,
\begin{eqnarray}\label{Jt}
 \rho = C(r) G^{r t}\,,\quad \partial_r \rho=0\,,
\end{eqnarray}
which describes the charge density of the dual boundary theory.

To construct the holographic DC conductivity matrix associated with the conserved current $J^\mu$ dual to the $U(1)$ gauge field $A_\mu$ we follow the prescription developed by~\cite{SBlake:2014yla}.
In particular, we consider the following set of perturbations to the background~\eqref{bk},
\begin{equation}
\begin{split}
\delta g_{ti} = C(r) h_{ti}(r), \hspace{.5cm} \delta  g_{ri} = C(r) h_{ri}\,, \\
\delta A_{i} = -E_i t + a_i(r), \hspace{.5cm} \delta \psi^1  = \chi_1(r), \hspace{.5cm} \delta \psi^2 = \chi_2(r)\,,
\end{split}
\end{equation}
with $i = x,y$. 
Maxwell's equations~\eqref{maxwell} along the radial direction are then of the form $\partial_r J^i = 0$, with $J^i=\sqrt{-g}\, G^{ir}$ the spatial components of the conserved current in the dual theory,
\begin{align}
\label{currentx}
J^x = -\mathcal{L}^{(1,0,0)}(s,p,\phi) \bigg(h U(r) h_{ry}(r) + C(r) h_{tx}(r)A_t'(r) + U(r) a_x'(r) \bigg)  - E_y \, \mathcal{L}^{(0,1,0)}(s,p,\phi) \, ,
\end{align}
\begin{align}
\label{currenty}
J^y = \mathcal{L}^{(1,0,0)}(s,p,\phi) \bigg(h U(r) h_{rx}(r) - C(r) h_{ty}(r)A_t'(r) - U(r) a_y'(r) \bigg) + E_x \, \mathcal{L}^{(0,1,0)}(s,p,\phi)\, ,
\end{align}
where we have defined 
\be
\mathcal{L}^{(1,0,0)} \equiv \frac{\partial \mathcal{L}(s,p,\phi)}{ \partial s} \, , \qquad  \mathcal{L}^{(0,1,0)} \equiv \frac{\partial \mathcal{L}(s,p,\phi)}{ \partial p} \, .
\ee 
Since the currents $(J^x, J^y)$ are radially conserved, they may be calculated anywhere within the bulk. The horizon is a convenient choice. 

Since the background geometry is regular near the horizon $r = r_h$, we have
\begin{equation}
\begin{split}
A_t =& A_t'(r_h)(r-r_h) +  \ldots \, , \\
U =& U'(r_h)(r-r_h) + \ldots = 4 \pi \, T(r-r_h) +  \ldots \, .
\end{split}
\end{equation}
Then the constraint of regularity of the perturbation equations imposes the near-horizon expansions
\begin{equation}
\begin{split}
a_i (r) =& -\frac{E_i}{4 \pi T} \log(r-r_h) + \ldots \, , \\  
h_{ti} (r) =& U(r) h_{ri}(r) + \ldots \, ,  \quad\quad \chi_i (r) = \chi_i(r_h) + \ldots \, .
\end{split}
\end{equation}
The horizon data for $h_{tx}$ and $h_{ty}$ can be extracted from the perturbed Einstein's equations~\eqref{einstein} using the regularity conditions, yielding
\begin{equation}
\begin{split}
\label{metricperts}
h_{tx}(r_h) &= \frac{\mathcal{L}^{(1,0,0)} (\text{C}^2 A_t' \left(E_x k^2 Y- E_y h A_t' \mathcal{L}^{(1,0,0)} \right) -E_y h^3 \mathcal{L}^{(1,0,0)}+E_y h k^2 \text{C} Y)}{\text{C}^2 \left(h^2 A_t'^2   \mathcal{L}^{(1,0,0)^2}+k^4 Y^2\right)-2 h^2 k^2 \text{C} Y \mathcal{L}^{(1,0,0)}+h^4 \mathcal{L}^{(1,0,0)^2}}\,,  \\
h_{ty}(r_h) &= \frac{\mathcal{L}^{(1,0,0)}(\text{C}^2 A_t' (E_x h A_t' \mathcal{L}^{(1,0,0)}+E_y k^2 Y) +E_x h^3 \mathcal{L}^{(1,0,0)}-E_x h k^2 \text{C} Y)}{\text{C}^2 \left(h^2 A_t'^2 \mathcal{L}^{(1,0,0)^2}+k^4 Y^2\right)-2 h^2 k^2 \text{C} Y \mathcal{L}^{(1,0,0)}+h^4 \mathcal{L}^{(1,0,0)^2}}  \, .
\end{split}
\end{equation}
From here on it should be understood that all functions are evaluated at the horizon $r=r_h$. 
Substituting the expressions (\ref{metricperts}) for the metric perturbations into (\ref{currentx}) and (\ref{currenty}) finally yields Ohm's law in matrix form,
\begin{eqnarray}
\label{Jmatrix}
 \left( \begin{array}{c}
 J^x \\ J^y 
\end{array} \right)  = \sigma \left( \begin{array}{c}
E_x \\ E_y 
\end{array} \right)  ,
\end{eqnarray}
with the components of the conductivity matrix $\sigma_{ij}$ given by 
\begin{equation}
\begin{split}
\sigma_{xx} =& \sigma_{yy} = \dfrac{k^2 C Y (k^2 C Y - h^2  \mathcal{L}^{(1,0,0)} - 
C^2 (A^\prime _t)^{2} \mathcal{L}^{(1,0,0)})}{h^2 (-2 k^2 C Y + h^2 \mathcal{L}^{(1,0,0)} + C^2 A_t'^2 \mathcal{L}^{(1,0,0)}) + C^2 k^4 Y^2 {\mathcal{L}^{(1,0,0)}}^{-1}}\, ,  \\
\sigma_{xy} =&-\sigma_{yx}=
\dfrac{h C A_t' \mathcal{L}^{(1,0,0)} (-2 k^2 C Y + h^2  \mathcal{L}^{(1,0,0)} + C^2 A_t'^2  \mathcal{L}^{(1,0,0)})}{h^2 (-2 k^2 C Y + h^2 \mathcal{L}^{(1,0,0)} + C^2 A_t'^2 \mathcal{L}^{(1,0,0)}) + C^2 k^4 Y^2 {\mathcal{L}^{(1,0,0)}}^{-1}} -\mathcal{L}^{(0,1,0)}\, . \\
\end{split}
\end{equation}
These can be written in a slightly more compact form by defining the quantities
\be
\xi \equiv k^2 C Y - \left(h^2  + C^2 A_t'^2 \right) \mathcal{L}^{(1,0,0)} \, ,  \quad
\Omega \equiv -(k^2 C Y + \xi) \, ,
\ee
in terms of which we have the expressions included in the main body of the paper,
\begin{equation}
\begin{split}\label{sigmas}
\sigma_{xx} =& \sigma_{yy} =  \dfrac{k^2 C Y \xi}{h^2 \Omega + C^2 k^4 Y^2 {\mathcal{L}^{(1,0,0)}}^{-1}} \,, \\
\sigma_{xy} =&-\sigma_{yx}= \dfrac{h C A_t' \, \mathcal{L}^{(1,0,0)} \, \Omega}{h^2 \Omega + C^2 k^4 Y^2 {\mathcal{L}^{(1,0,0)}}^{-1}} - \mathcal{L}^{(0,1,0)} \, .
\end{split}
\end{equation}
Notice that at this stage they depend on the gauge field term $A_t'$ and not on the charge density $\rho$. 
Since the relationship between $A_t'$ and $\rho$  in these theories is generically nonlinear, we will be able to express (\ref{sigmas})  explicitly in terms of the charge density only in special cases.
Also, while both conductivities depend on the standard Maxwell term $s$ through the dependence on 
$ \mathcal{L}^{(1,0,0)}$, only $\sigma_{xy}$ is sensitive to $\mathcal{L}^{(0,1,0)}$.

As a simple check of our analysis we consider the standard EMD theory described by 
\be
\mathcal{L}(s,p,\phi) = Z(\phi)s \, , 
\ee 
for which $\mathcal{L}^{(1,0,0)} = Z$ and $\mathcal{L}^{(0,1,0)} = 0$. 
Expressed in terms of the charge density $\rho = C Z A_t'$,
the corresponding conductivities (\ref{sigmas}) are of the form
\begin{eqnarray}
\label{AppSigmas}
\sigma_{xx} &=&  \sigma_{yy} = \dfrac{k^2 C Y (k^2 C Y Z - h^2  Z^2 - \rho^2)}{h^2 (-2 k^2 C Y Z + h^2 Z^2 + \rho^2) + C^2 k^4 Y^2 } \, , \nn \\
\sigma_{xy} &=& \sigma_{yx} = 
\dfrac{h \rho (-2 k^2 C Y Z + h^2  Z^2 + \rho^2)}{h^2 (-2 k^2 C Y Z + h^2 Z^2 + \rho^2) + C^2 k^4 Y^2 } \, , 
\end{eqnarray}
in agreement with the results of \cite{SBlake:2014yla}, as expected.
In addition, for the special case without a scalar sector, the quantities (\ref{sigmas}) agree with those computed in 
\cite{SWang:2018hwg}, a nontrivial check on our analysis. 

Finally, from the conductivity matrix we can extract the inverse Hall angle,
\begin{eqnarray}
\cot \Theta_H = \dfrac{\sigma_{xx}}{\sigma_{xy}}\,,
\end{eqnarray}
and the resistivity matrix,
\begin{eqnarray}
R_{xx} = R_{yy} = \dfrac{\sigma_{xx}}{\sigma_{xx}^2 + \sigma_{xy}^2}\,, \hspace{.5cm} R_{xy} = -R_{yx} = -\dfrac{\sigma_{xy}}{\sigma_{xx}^2 + \sigma_{xy}^2}\,.
\end{eqnarray}
Written in terms of $\xi$ and $\Omega$ we have
\begin{eqnarray}
\label{AppGen}
&& \cot \Theta_H = \dfrac{k^2 \xi C Y}{(h^2 \Omega + k^4 C^2 Y^2 {\mathcal{L}^{(1,0,0)}}^{-1})\mathcal{L}^{(0,1,0)} - h C \Omega A_t' \mathcal{L}^{(1,0,0)}}, \nn \\
&& R_{xx} = \dfrac{k^2 \xi C Y}{\big(h^2 \Omega + k^4 C^2 Y^2 {\mathcal{L}^{(1,0,0)}}^{-1} \big) \bigg[\frac{k^4 \xi^2 C^2 Y^2 }{(h^2 \Omega + k^4 C^2 Y^2 {\mathcal{L}^{(1,0,0)}}^{-1})^2} + \bigg(\mathcal{L}^{(0,1,0)} - \frac{h C A_t' \Omega {\mathcal{L}^{(1,0,0)}}^2}{(h^2 \Omega \mathcal{L}^{(1,0,0)}  + k^4 C^2 Y^2)} \bigg)^2 \bigg]} \,, \nn \\
&& R_{xy} = \dfrac{\dfrac{h C A_t' \mathcal{L}^{(1,0,0)} \Omega}{h^2 \Omega + C^2 k^4 Y^2 {\mathcal{L}^{(1,0,0)}}^{-1}} - \mathcal{L}^{(0,1,0)}}{ \bigg[\frac{k^4 \xi^2 C^2 Y^2 }{(h^2 \Omega + k^4 C^2 Y^2 {\mathcal{L}^{(1,0,0)}}^{-1})^2} + \bigg(\mathcal{L}^{(0,1,0)} - \frac{h C A_t' \Omega {\mathcal{L}^{(1,0,0)}}^2}{(h^2 \Omega \mathcal{L}^{(1,0,0)}  + k^4 C^2 Y^2)} \bigg)^2 \bigg]}.
\end{eqnarray}
These expressions are entirely general and describe the conductivities resulting from theories of the form (\ref{action}). 
Once the background solution~\eqref{bk} is known, $A_t'(r_h)$ and $r_h$ can be expressed in terms of $(T, \rho, h, k)$ by solving~\eqref{temp} and~\eqref{Jt}. Therefore, the resistivity and the Hall angle are general functions of the temperature $T$, the charge density $\rho$, the magnetic field $h$ and the strength of momentum dissipation $k$.

We close by elaborating on the connection between the probe limit and the limit in which momentum dissipation is the dominant physical scale. 
Computing from scratch the conductivities under the assumption that the gauge field sector $\mathcal{L_{\text{U(1)}}}=\mathcal{L}(s,p,\phi)$ is a probe and does not backreact on the geometry, 
which can be done by sending 
$\mathcal{L_{\text{U(1)}}} \rightarrow \delta \mathcal{L_{\text{U(1)}}}$ with $\delta$ a perturbatively small parameter,
yields the expansions
\begin{eqnarray}
\label{Exp}
&& \sigma_{xx} \approx \mathcal{L}^{(1,0,0)}  + \delta \, \dfrac{(h^2 - C^2 A_t'^2) \, {\mathcal{L}^{(1,0,0)}}^2 }{C Y k^2}  + \mathcal{O}(\delta^2) , \nn \\
&& \sigma_{xy} \approx -\mathcal{L}^{(0,1,0)}  -  \delta \,  \dfrac{2 h A_t' \, {\mathcal{L}^{(1,0,0)}}^2 }{Y k^2} + \mathcal{O}(\delta^2) \, ,
\end{eqnarray}
from which we extract the probe limit result we have used in the main body of the paper,
\begin{eqnarray}
R_{xx} \approx \dfrac{\mathcal{L}^{(1,0,0)}}{{\mathcal{L}^{(0,1,0)}}^2 + {\mathcal{L}^{(1,0,0)}}^2}\, , \qquad \cot \Theta_H \approx  -\dfrac{\mathcal{L}^{(1,0,0)}}{\mathcal{L}^{(0,1,0)}} \, .
\end{eqnarray}
On the other hand, if we expand the general expressions (\ref{AppSigmas}) in a large momentum dissipation expansion, by sending $k^2 \rightarrow \frac{k^2}{\delta}$ with $\delta \rightarrow 0$, 
we obtain the same result (\ref{Exp}).
Thus, from the point of view of the holographic conductivities, working under the assumption that the charge degrees of freedom are a probe is equivalent to assuming that 
the momentum dissipation scale $k$ dominates over the other physical scales in the system, \emph{i.e.} the charge density $\rho$ and magnetic field $h$.
We stress that this is not necessarily a large-$k$ limit, but rather a statement about the hierarchy between $k$, $\rho$ and $h$.

\vspace{0.2in}
\qquad\qquad\qquad\qquad\qquad\qquad\qquad\qquad{\textbf{Magnetotransport in the DBI Theory}}
\vspace{0.1in}

In this section we include the magnetotransport results for a theory whose gauge field sector is described by the nonlinear DBI model, 
\begin{equation}\label{dbiaction}
S_{DBI}=\sqrt{-g}-\sqrt{-\det(g+Z^{1/2}\, F)}=\sqrt{-g}\left[1-\sqrt{1-2(Z s + Z^2 p^2/2})\right]\,,
\end{equation}
and which can be used to describe an ensemble of probe charge carriers interacting 
with a larger neutral quantum critical bath. 
Note that (\ref{dbiaction}) corresponds to the special case $Z_1(\phi) = 1$, $Z_2(\phi) =  \sqrt{Z (\phi)}$ of the theory studied in \cite{SCremonini:2017qwq,SBlauvelt:2017koq,SKiritsis:2016cpm}.

The associated in-plane resistivity and Hall angle are given by
\begin{equation}
R_{xx}=\frac{C}{\sqrt{Z}}\frac{\sqrt{\rho^2+Z(C^2+h^2 Z)}}{\rho^2+C^2 Z},\qquad \cot\Theta=\frac{C}{h\rho\sqrt{Z}}\sqrt{\rho^2+Z(C^2+h^2 Z)}\,,
\end{equation}
with $\rho$ denoting the charge density.
Following the discussion in the main text, we introduce two constants $z_0$ and $c_0$ through $Z=z_0/T$ and $C=c_0 T$, and construct the dimensionless expressions
\begin{equation}\label{rescale}
\textbf{T}=\frac{c_0^2 z_0}{\rho^2}T,\qquad \textbf{h}=\frac{c_0^2 z_0^2}{\rho^3}h\,.
\end{equation}
We then have 
\begin{equation}
\begin{split}
R_{xx}&=\zeta_0 \frac{\mathbf{T}^{3/2}}{1+\mathbf{T}} \sqrt{1+\mathbf{T}+\mathbf{h}^2/\mathbf{T}^2} ,\\
\cot\Theta_H&=\frac{\mathbf{T}^{3/2}}{\mathbf{h}}\sqrt{1+\mathbf{T}+\mathbf{h}^2/\mathbf{T}^2}\,,\\
MR&=\sqrt{1+\frac{\mathbf{h}^2}{\mathbf{T}^2(1+\mathbf{T})}}-1\,.
\end{split}
\end{equation}
In the ``high-temperature'' limit $\mathbf{T}\gg 1+\mathbf{h}^2/\mathbf{T}^2$, 
one obtains \cite{Blauvelt:2017koq} the simple scalings
\begin{equation}
R_{xx}\sim\zeta_0 \textbf{T}, \;\; \cot \Theta_H\sim\frac{\textbf{T}^2}{\textbf{h}}\,,
\end{equation}
while the magnetoresistance reads
\begin{equation}
MR\sim \frac{\textbf{h}^2}{\textbf{T}^3}\,.
\end{equation}
It is now obvious that the DBI theory~\eqref{dbiaction} gives a weak-field magnetoresistance that goes as $h^2/T^3$, instead of the observed $h^2/T^4$ for the cuprates, and therefore it could be a contender to describe strange metals other than the cuprates.
Finally, when $\textbf{h}$ dominates over the temperature we obtain the strong-field magnetotransport results,
\begin{equation}
R_{xx}\sim\zeta_0 \frac{\mathbf{T}^{1/2}}{1+\mathbf{T}}\, \mathbf{h}, \;\; \cot \Theta_H\sim\mathbf{T}^{1/2},\;\; MR\sim \frac{\mathbf{h}}{\mathbf{T}\sqrt{1+\mathbf{T}}}\,.
\end{equation}
We find that both $R_{xx}$ and $MR$ scale linearly with magnetic field, while the Hall angle $\cot \Theta_H$ is almost $\mathbf{h}$ independent. 


\end{document}